\documentclass[conference]{IEEEtran}

\newcommand{\e}{{\rm e}}
\usepackage{amsfonts}
\usepackage{color}
\usepackage{amsmath}
\usepackage{amsthm}
\usepackage{amssymb}
\usepackage{t1enc}
\usepackage{alltt}
\usepackage{epsfig}
\usepackage{mathrsfs}
\usepackage{graphicx,ifthen}
\usepackage[scaled]{helvet}

\makeatletter

\newcommand{\Rmnum}[1]{\expandafter\@slowromancap\romannumeral #1@}

\begin{document}

\title{On the Exact Distribution of the Scaled Largest Eigenvalue}
\author{
\IEEEauthorblockN{Lu Wei and Olav Tirkkonen}
\IEEEauthorblockA{Department of Communications and Networking, \\
Aalto University\\
P.O. Box 13000, Aalto-00076, Finland \\
Email: \{lu.wei, olav.tirkkonen\}@aalto.fi}
\and
\IEEEauthorblockN{Prathapasinghe Dharmawansa and Matthew McKay}
\IEEEauthorblockA{Department of Electronic and Computer Engineering, \\
Hong Kong University of Science and Technology\\
Clear Water Bay, Kowloon, Hong Kong \\
Email: \{eesinghe, eemckay\}@ust.hk}
}
\maketitle

\begin{abstract}
In this paper we study the distribution of the scaled largest
eigenvalue of complex Wishart matrices, which has diverse applications
both in statistics and wireless communications. Exact expressions,
valid for any matrix dimensions, have been derived for the probability
density function and the cumulative distribution function. The derived
results involve only finite sums of polynomials. These results are
obtained by taking advantage of properties of the Mellin transform for
products of independent random variables.
\end{abstract}
\


\IEEEpeerreviewmaketitle

\section{Introduction}\label{sec:intro}

% eigenstats, motivation
Eigenvalue statistics of Wishart matrices play a key role in the performance analysis and design of various communication systems. Among these, the distribution of Scaled Largest Eigenvalue (SLE), defined as the ratio of the largest eigenvalue to the normalized sum of all eigenvalues, has been shown to be an important measure. The applicability of the SLE spans from classical problems in statistics~\cite{1972Johnson,1972Davis,1973Schurmann,1974Krishnaiah} to modern applications in wireless communications~\cite{2006Besson,2008bZeng,2010Bianchi,2010Boaz,2011LuA}. Classical problems include testing the presence of interactions in a two-way model~\cite{1972Johnson} and testing the equality of eigenvalues of certain matrices against various of alternatives~\cite{1972Davis,1973Schurmann,1974Krishnaiah}. Contemporary applications in wireless communications include non-parametric detection in array processing~\cite{2006Besson} and spectrum sensing in cognitive radio networks~\cite{2008bZeng,2010Bianchi,2010Boaz,2011LuA}. Specifically, for spectrum sensing applications, the SLE is formulated as a test statistics, which is first proposed by~\cite{2008bZeng} and further investigated in~\cite{2010Bianchi,2010Boaz,2011LuA}. The SLE based detector is the best known detector for single source detection, outperforming several classical detectors in realistic sensing
scenarios~\cite{2008bZeng,2010Bianchi,2011LuA}. Despite the importance of the knowledge of the SLE, existing results on its statistical properties are rather limited. In this paper, we aim to address this problem by deriving simple and exact expressions for the Probability Density Function (PDF) and Cumulative Distribution Function (CDF) of the SLE.

The rest of this paper is organized as follows. In Section~\ref{sec:def} we formally define the scaled largest eigenvalue of Wishart matrix followed by a concise survey on existing results. Section~\ref{sec:main} is devoted to deriving the exact SLE distribution as well as the closed-form coefficients. Numerical examples are provided in~\ref{sec:simu} to verify the derived results. Finally in Section~\ref{sec:conc} we conclude main results of this paper.

\section{Definitions, Prior Results and Contributions}\label{sec:def}

Define a $K\times N$ ($K\leq N$) dimensional random matrix
$\mathbf{X}$ with independent and identically distributed (i.i.d)
complex Gaussian entries, each with zero mean and unit variance. The
$K \times K$ Hermitian matrix\footnote{The operator $(\cdot)^\dag$ denotes conjugate-transpose.}
$\mathbf{R}=\mathbf{X}\mathbf{X}^{\dag}$ follows a complex Wishart
distribution with $N$ degrees of freedom (d.o.f). We denote the
ordered eigenvalues of $\mathbf{R}$ as
$\lambda_{1}>\lambda_{2}>...>\lambda_{K}>0$, and the normalized trace
of $\mathbf{R}$ as
$T=\text{tr}\{\mathbf{R}\}/K=\left(\sum_{i=1}^{K}\lambda_{i}\right)/K$.
The scaled largest eigenvalue of $\mathbf{R}$ is formally defined as
the ratio of its largest eigenvalue to its normalized trace, i.e.,
\begin{equation}
X :=\frac{\lambda_{1}}{\frac{1}{K}\sum_{i=1}^{K}\lambda_{i}}=\frac{\lambda_{1}}{T},
\end{equation}
where it can be verified that $x\in[1,K]$.

\begin{figure}[t!]
\centering
\includegraphics[width=3.5in]{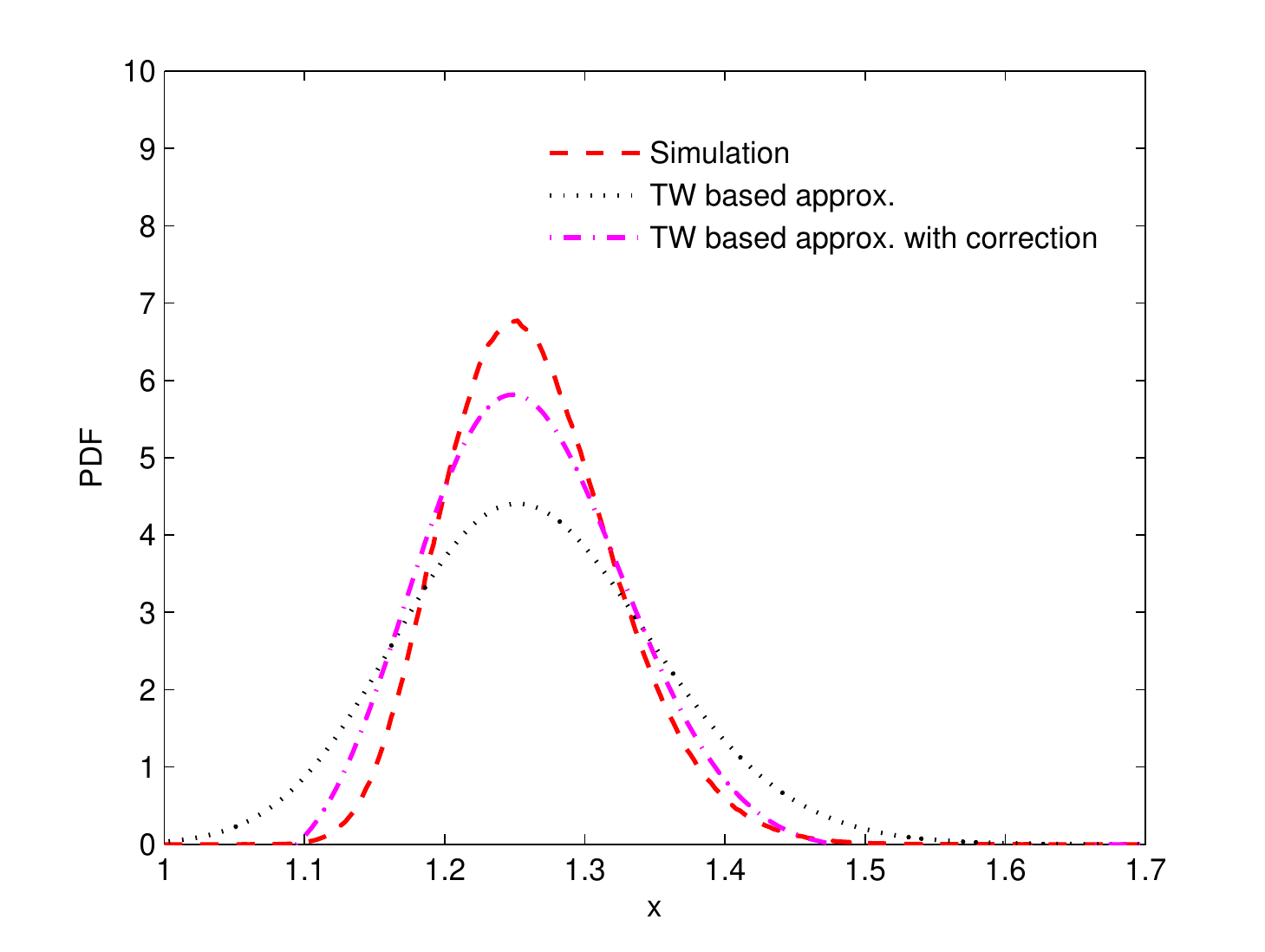}
\caption{Accuracy of some existing asymptotic approximations to the SLE distribution: $K=4$, $N=100$. The `TW based approx.' refers to the result of~\cite{2010Bianchi} and the `TW based approx. with correction' refers to the result of~\cite{2010Boaz}.}\label{fig:1}
\end{figure}

The distribution of $X$ has been the subject of intense study in the
literature. An exact expression for the distribution of $X$ in terms of a high dimensional integral has been proposed in~\cite{1972Johnson}.
In~\cite{1972Davis}, a relation between Laplace transforms of random variables $X$ and $\lambda_{1}$ was established. By symbolically inverting the Laplace transforms, some representations for the distribution of $X$ were derived in~\cite{1973Schurmann,1974Krishnaiah}. Whilst being exact, these representations~\cite{1972Johnson,1972Davis,1973Schurmann,1974Krishnaiah} can only be evaluated numerically for small values of $K$ and $N$ due to their unexplicit and complicated forms. Recently, motivated by its application in spectrum sensing, several
asymptotical\footnote{Asymptotic in the sense that the matrix
dimensions go to infinity while their ratio is kept fixed, i.e.
$K\rightarrow\infty$, $N\rightarrow\infty$ and $K/N\rightarrow
r\in(0,1)$.} distributions of $X$ have been
derived~\cite{2010Boaz,2010Bianchi,2011LuA} via random matrix theory.
Although these results are easy to compute, their accuracy can not be
guaranteed for not-so-large $K$ and $N$. As an example, in
Fig.~\ref{fig:1}, we illustrate the accuracy of an asymptotic result
based on Tracy-Widom distribution (`TW based approx.')
from~\cite{2010Bianchi} and an improved version (`TW based approx.
with correction') from~\cite{2010Boaz} with a typical choice of
parameters in spectrum sensing\footnote{Corresponding to a situation of a sensing device with
$4$ antennas with $100$ samples per antenna.}: $K=4$ and
$N=100$. From Fig.~\ref{fig:1} we can see that the approximation errors of both asymptotic results
are non-negligible. Note that for other applications discussed in
Section~\ref{sec:intro}, the $K$, $N$ values are often smaller than
in spectrum sensing applications. For example,
in~\cite{1972Johnson} the authors considered $K\leq4$ and $N\leq100$
in all the simulations. In addition, for the application considered
in~\cite{2006Besson} it was remarked that choosing $K>2$ does not result in any performance improvement and there $N$ was always chosen to be no more than $20$. Therefore, the approximation accuracy may become even lower when using the existing asymptotic results for the above applications.

In this work, we derive exact expressions for the SLE distribution,
valid for arbitrary $K$ and $N$, as a finite sum of polynomials with some unknown coefficients. Closed-form expressions for the coefficients are  obtained for the most useful system configurations of $K=2, 3, 4$ with any $N$. These results are simple to calculate and do not involve any integral representations. To obtain these results, we adopt a novel approach based on the Mellin transform, which eliminates the need to handle correlations between $\lambda_{1}$ and $T$. The derived results yield a useful analytical tool in applications involving statistical properties of the SLE.

\section{The SLE Distribution}\label{sec:main}

\subsection{Exact Distribution}

Although there exists intractable correlation between random variables $\lambda_{1}$ and $T$, it has been proved in~\cite{2006Besson} that the random variables $X$ and $T$ are independent. As such, $\lambda_{1}$ equals the product of the independent random variables $X$ and $T$. By this independence, the $(z-1)$th moment of $\lambda_{1}$ can be represented as
\begin{equation}
E[\lambda_{1}^{z-1}]=E[(XT)^{z-1}]=E[X^{z-1}]E[T^{z-1}].
\end{equation}
Moreover, the $(z-1)$th moment of a random variable $x$, with PDF
$p(x)$, equals its Mellin transform as
\begin{equation}
E[x^{z-1}]=\int_{0}^{\infty}x^{z-1}p(x)\mathrm{d}x:=\mathcal{M}_{z}[p(x)],
\end{equation}
where $\mathcal{M}_{z}[\cdot]$ denotes the Mellin transform operation. Define
$f_{\lambda_{1}}(x)$, $f_{T}(x)$ and $f_{X}(x)$ as the PDFs of
$\lambda_{1}$, $T$ and $X$ respectively. We have
\begin{equation}\label{eq:MelPro}
\mathcal{M}_{z}[f_{\lambda_{1}}(x)]=\mathcal{M}_{z}[f_{T}(x)]\mathcal{M}_{z}[f_{X}(x)].
\end{equation}
By the Mellin inversion theorem, the PDF of $X$ can be
uniquely determined by the following contour integral
\begin{equation}\label{eq:InvMellin}
f_{X}(x)=\frac{1}{2\pi i}\int_{c-i\infty}^{c+i\infty}x^{-z}\frac{\mathcal{M}_{z}[f_{\lambda_{1}}(x)]}{\mathcal{M}_{z}[f_{T}(x)]}\mathrm{d}z.
\end{equation}
In principle, the above Mellin inversion integral can be evaluated by using the residue theorem. Note that, following the above Mellin transform framework, a related distribution of the trace to the smallest eigenvalue has been derived recently~\cite{2011Lu}.

The PDF of $\lambda_{1}$ admits the following representation~\cite{2003Dighe,2005Maaref},
\begin{equation}\label{eq:lamPDF}
f_{\lambda_{1}}(x)=\sum_{i=1}^{K}\e^{-ix}\sum_{j=N-K}^{(N+K)i-2i^{2}}c_{i,j}x^{j},
\end{equation}
where $c_{i,j}$ denotes the unknown coefficients. Closed-form coefficients formulas will be derived in the next subsection. Meanwhile, numerical algorithms are also available in~\cite{2003Dighe,2005Maaref} to calculate $c_{i,j}$ for a given $K$ and $N$.

In order to apply the Mellin transform framework, we first need to calculate $\mathcal{M}_{z}[f_{\lambda_{1}}(x)]$, which equals
\begin{equation}\label{eq:Melfx}
\mathcal{M}_{z}[f_{\lambda_{1}}(x)]=\sum_{i=1}^{K}\sum_{j=N-K}^{(N+K)i-2i^{2}}\frac{c_{i,j}}{i^{j}}\frac{\Gamma(z+j)}{i^{z}}.
\end{equation}
It is well known that the sum of all eigenvalues of $\mathbf{R}$,
$\sum_{i=1}^{K}\lambda_{i}$, follows central Chi-square distribution
with $2KN$ degrees of freedom, therefore the PDF of
$T=\left(\sum_{i=1}^{K}\lambda_{i}\right)/K$ can be obtained as
\begin{equation}
f_{T}(x)=\frac{K^{KN}}{(KN-1)!}x^{KN-1}\e^{-Kx}.
\end{equation}
Its Mellin transform is
\begin{equation}\label{eq:Melgx}
\mathcal{M}_{z}[f_{T}(x)]=\frac{K^{1-z}}{(KN-1)!}\Gamma(z+KN-1).
\end{equation}

Inserting~(\ref{eq:Melfx}) and~(\ref{eq:Melgx}) into the Mellin inversion integral~(\ref{eq:InvMellin}) we have
\begin{equation}\label{eq:fX}
f_{X}(x)=\frac{(KN-1)!}{K}\sum_{i=1}^{K}\sum_{j=N-K}^{(N+K)i-2i^{2}}\frac{c_{i,j}}{i^{j}}A(x,z)
\end{equation}
where
\begin{equation}
A(x,z)=\frac{1}{2 \pi i}\int_{c-i\infty}^{c+i\infty}\frac{\Gamma(z+j)}{\Gamma(z+KN-1)}\left(\frac{ix}{K}\right)^{-z}\mathrm{d}z.
\end{equation}

\begin{figure*}[!t]
\hrulefill\vspace*{4pt}
\begin{equation}\label{eq:MeijerG}
G_{p,q}^{m,n}\left( x \left|
\begin{array} {c}
a_{1},\ldots,a_{p} \\ b_{1},\ldots,b_{q} \\
\end{array} \right.\right) = \frac
1{2\pi i} \int_{c-i\infty}^{c+i\infty}{\frac{\prod_{j=1}^m
\Gamma(b_j+z) \prod_{j=1}^n \Gamma\left(1-a_j-z\right)}
{\prod_{j=n+1}^p \Gamma(a_{j}+s) \prod_{j=m+1}^q
\Gamma\left(1-b_j-z\right)}} x^{-z} \,\mathrm{d}z.
\end{equation}
\hrulefill\vspace*{4pt}

\begin{equation}\label{eq:fXsimp}
f_{X}(x)=\frac{(KN-1)!}{K^{KN-1}}\sum_{i=1}^{K}\sum_{j=N-K}^{(N+K)i-2i^{2}}\frac{i^{KN-j-2}}{(KN-j-2)!}c_{i,j}x^{j}\left(\frac{K}{i}-x\right)^{KN-j-2}\theta\left(1-\frac{ix}{K}\right).
\end{equation}
\hrulefill\vspace*{4pt}

\begin{equation}\label{eq:FXsimp}
F_{X}(y)=\frac{(KN-1)!}{K^{KN-1}}\sum_{i=1}^{K}\sum_{j=N-K}^{(N+K)i-2i^{2}}i^{KN-j-2}c_{i,j}\left(C(y)\theta\left(\frac{K}{i}-y\right)+C\left(\frac{K}{i}\right)\theta\left(y-\frac{K}{i}\right)-C(1)\right),
\end{equation}
where
\begin{equation*}
C(y)=\left(\frac{K}{i}\right)^{KN-j-2}\sum_{q=0}^{KN-j-1}\frac{(-i/K)^{q}(j+q+1)^{-1}}{(KN-j-2-q)!q!}y^{q+j+1}.
\end{equation*}
\hrulefill\vspace*{4pt}
\end{figure*}

By definition of the Meijer's G function~\cite{1990Prudnikov}, as shown in~(\ref{eq:MeijerG}) on top of this page, the function $A(x,z)$ can now be represented as
\begin{equation}\label{eq:Meijer}
A(x,z)=G_{1,1}^{1,0}
\left(\frac{ix}{K}\left|
\begin{array}{c}
KN-1 \\ j \\
\end{array}
\right.\right).
\end{equation}
By using the fact that
\begin{equation}
G_{1,1}^{1,0}
\left(x\left|
\begin{array}{c}
a \\ b \\
\end{array}
\right.\right)=\frac{x^{b}(1-x)^{a-b-1}}{(a-b-1)!}\theta(1-x),
\end{equation}
where $\theta(\cdot)$ denotes the Heaviside step function
\begin{equation}
\theta(x)= \left\{
  \begin{array}{l l}
     0 & \quad x<0\\
     1 & \quad x\geq 0\\
   \end{array} \right.,
\end{equation}
the PDF of $X$ in (\ref{eq:fX}) simplifies to the expression shown in~(\ref{eq:fXsimp}) on top of this page.

We now focus on the CDF. By definition, the CDF of $X$, equals
\begin{equation}\label{eq:CDF}
F_{X}(y)=\frac{\Gamma(KN)}{K^{KN-1}}\sum_{i=1}^{K}\sum_{j=N-K}^{(N+K)i-2i^{2}}i^{KN-j-2}c_{i,j}B(y),
\end{equation}
where
\begin{equation}
B(y)=\frac{\int_{1}^{y}x^{j}\left(\frac{K}{i}-x\right)^{KN-j-2}\theta\left(1-\frac{ix}{K}\right)\mathrm{d}x}{(KN-j-2)!}
\end{equation}
and $y\in[1,\infty)$. Using the definition of the hypergeometric function
$_{2}F_{1}(a,b;c;x)=\sum_{n=0}^{\infty}\frac{(a)_{n}(b)_{n}}{(c)_{n}}\frac{x^{n}}{n!},$
where $(a)_{n}=\Gamma(a+n)/\Gamma(a)$ defines the Pochhammer symbol,
the function $B(y)$ becomes
\begin{equation}\label{eq:Bsimp}
B(y)=C(y)\theta\left(\frac{K}{i}-y\right)+C\left(\frac{K}{i}\right)\theta\left(y-\frac{K}{i}\right)-C(1),
\end{equation}
where
\begin{eqnarray}
C(y)&=&\frac{(K/i)^{KN-j-2}}{(KN-j-2)!(j+1)}y^{j+1}\times\nonumber\\
&&_{2}F_{1}\left(-KN+j+2,j+1;j+2;\frac{i}{K}y\right).
\end{eqnarray}
Since the parameters $a$, $b$ and $c$ of take integer values, the
hypergeometric function can be simplified to a \emph{finite} sum of polynomials. Consequently the function $C(y)$, after some manipulations, equals
\begin{equation}\label{eq:Csimp}
C(y)=\left(\frac{K}{i}\right)^{KN-j-2}\sum_{q=0}^{KN-j-1}\frac{(-i/K)^{q}(j+q+1)^{-1}}{(KN-j-2-q)!q!}y^{q+j+1}.
\end{equation}
Inserting~(\ref{eq:Bsimp}) into~(\ref{eq:CDF}), the CDF expression is summarized as~(\ref{eq:FXsimp}) on top of this page.

Note that both the exact PDF expression~(\ref{eq:fXsimp}) and CDF expression~(\ref{eq:FXsimp}) involve the unknown coefficients $c_{i,j}$ inherited from~(\ref{eq:lamPDF}). Closed-form expressions for the coefficients $c_{i,j}$ will be derived for $K\leq4$ with arbitrary $N$. For other $K$ values, one has to resort to numerical techniques~\cite{2003Dighe,2005Maaref} to obtain the values of $c_{i,j}$.

\subsection{Closed-form Coefficients}
In order to circumvent possible computational burden when using the numerical algorithms~\cite{2003Dighe,2005Maaref} to compute $c_{i,j}$,
here we derive closed-form $c_{i,j}$ expressions for $K$ up to four with arbitrary $N$. Note that the considered cases $K\leq4$ cover the typical situations in  applications discussed in Section~\ref{sec:intro}.

%\newcounter{MYtempeqncnt}
\begin{figure*}[!t]
\hrulefill\vspace*{4pt}
% f_{\lambda_{1}}(x)=\e^{-x}\sum_{j=N-2}^{N}c_{1,j}x^{j}+\e^{-2x}\sum_{j=N-2}^{2N-4}c_{2,j}x^{j},
\begin{equation*}
K=2:~~~~
\end{equation*}
\begin{equation}\label{eq:coK2}
c_{1,j}=\frac{2(-1)^{j-N}(2N-j-2)!}{(-N+j+2)!(N-j)!(N-2)!(N-1)!},~~~
c_{2,j}=\frac{-(2N-j-2)(2N-j-3)}{(-N+j+2)!(N-1)!}.
\end{equation}
\hrulefill\vspace*{4pt}

% f_{\lambda_{1}}(x)=\e^{-x}\sum_{j=N-3}^{N+1}c_{1,j}x^{j}+\e^{-2x}\sum_{j=N-3}^{2N-2}c_{2,j}x^{j}+\e^{-3x}\sum_{j=N-3}^{3N-9}c_{3,j}x^{j},
\begin{equation*}
K=3:~~~~
\end{equation*}
\begin{eqnarray}
c_{1,j}&=&\sum_{k=\text{max}\{0,j-N+1\}}^{\text{min}\{j-N+3,2\}}\frac{2(-1)^{j-N+1}(N-k)!(2N-j+k-4)!(-N+j-2k+4)}{(2-k)!k!(N-j+k-1)!(-N+j-k+3)!(N-3)!(N-2)!(N-1)!}\label{eq:1coK3},\\
c_{2,j}&=&\sum_{k=\text{max}\{0,j-2N+4\}}^{\text{min}\{j-N+3,2\}}\frac{(-1)^{k}}{(2-k)!k!(-N+j-k+3)!}\Bigg(\frac{2(N-k)!(2N-j+k-3)!}{(2N-j+k-5)!(N-3)!(N-1)!}-\nonumber\\
&&\frac{(N-k-1)!(2N-j+k-2)!}{(2N-j+k-4)!(N-3)!(N-2)!}-\frac{(N-k+1)!(2N-j+k-4)!}{(2N-j+k-6)!(N-2)!(N-1)!}\Bigg),\label{eq:2coK3}\\
c_{3,j}&=&\sum_{k=\text{max}\{0,j-2N+5\}}^{\text{min}\{j-N+3,N-1\}}\frac{1}{2k!(-N+j-k+3)!(N-2)!}\Bigg(\frac{(N-k+1)!(2N-j+k-4)!}{(N-k-1)!(2N-j+k-6)!}-\nonumber\\
&&\frac{(N-k)!(2N-j+k-3)!(N-2)}{(N-k-2)!(2N-j+k-5)!(N-1)}\Bigg).\label{eq:3coK3}
\end{eqnarray}
\hrulefill\vspace*{4pt}

\end{figure*}

Following the methodology of obtaining the coefficients for the
smallest eigenvalue distribution~\cite{2008C.S.Park}, we first write
an integral representation for the largest eigenvalue distribution;
\begin{eqnarray}
f_{\lambda_{1}}(x)&=&\frac{D(K,N)}{(K-1)!}x^{N-K}\e^{-x}\int_{J}\prod_{2\leq i\leq j\leq K}(\lambda_{i}-\lambda_{j})^{2}\times\nonumber\\
&&\prod_{i=2}^{K}\lambda_{i}^{N-K}\e^{-\lambda_{i}}(x-\lambda_{i})^{2}\mathrm{d}\lambda_{i}\label{eq:larINT},
\end{eqnarray}
where the constant $D(K,N)=\left(\prod_{i=1}^{K}(N-i)!(K-i)!\right)^{-1}$ and the domain of the integration $J=[0,x]^{K-1}$. Similar to the case of the smallest eigenvalue~\cite{2008C.S.Park}, we first define the following integral
\begin{equation}
L_{a}(x):=\int_{0}^{x}\lambda^{a}(x-\lambda)^{2}\e^{-\lambda}\mathrm{d}\lambda,
\end{equation}
which, by repeated use of integration by parts, equals
\begin{equation}
L_{a}(x)=\sum_{k=0}^{2}\frac{2(a-k+2)!}{(-1)^{k}k!(2-k)!}x^{k}-\e^{-x}a!\sum_{k=0}^{a}\frac{(a-k+2)!}{k!(a-k)!}x^{k}.
\end{equation}

% may put double column eqs. here
When $K=2$, the distribution in~(\ref{eq:larINT}) becomes
\begin{equation}\label{eq:K2}
f_{\lambda_{1}}(x)=D(2,N)x^{N-2}\e^{-x}L_{N-2}(x).
\end{equation}
Comparing~(\ref{eq:K2}) with~(\ref{eq:lamPDF}) and after some
manipulations, the coefficients are obtained as~(\ref{eq:coK2}) on top of this page.

For $K=3$, it can be verified that~(\ref{eq:larINT}) equals
\begin{eqnarray}\label{eq:K3}
f_{\lambda_{1}}(x)=D(3,N)x^{N-3}\e^{-x}\bigg(L_{N-1}(x)L_{N-3}(x)-\nonumber\\
(L_{N-2}(x))^{2}\bigg).
\end{eqnarray}
By using the equality
\begin{equation}
\sum_{i=0}^{a}p_{i}x^{i}\sum_{i=0}^{b}q_{i}x^{i}=\sum_{i=0}^{a+b}\sum_{k=\text{max}\{0,i-b\}}^{\text{min}\{i,a\}}p_{k}q_{i-k}x^{i},
\end{equation}
and comparing~(\ref{eq:K3}) with~(\ref{eq:lamPDF}) the coefficient expressions can be calculated as shown in~(\ref{eq:1coK3})-(\ref{eq:3coK3}) on top of this page.

For $K=4$, equation~(\ref{eq:larINT}) can now be represented as
\begin{eqnarray}
f_{\lambda_{1}}(x)\!\!\!&=\!\!\!&D(4,N)x^{N-4}\e^{-x}\bigg(2L_{N-1}(x)L_{N-2}(x)L_{N-3}(x)+\nonumber\\
&&L_{N}(x)L_{N-2}(x)L_{N-4}(x)-(L_{N-3}(x))^{2}L_{N}(x)-\nonumber\\
&&(L_{N-1}(x))^{2}L_{N-4}(x)-(L_{N-2}(x))^{3}\bigg).\label{eq:K4}
\end{eqnarray}
Using the equality
\begin{eqnarray}
&&\sum_{i=0}^{a}p_{i}x^{i}\sum_{i=0}^{b}q_{i}x^{i}\sum_{i=0}^{c}l_{i}x^{i}=\nonumber\\
&&\sum_{i=0}^{a+b+c}\sum_{t=\text{max}\{0,i-c\}}^{\text{min}\{i,a+b\}}\sum_{k=\text{max}\{0,t-b\}}^{\text{min}\{t,a\}}p_{k}q_{t-k}l_{i-t}x^{i},
\end{eqnarray}
the closed-form coefficients for $K=4$ can be similarly obtained. They
are, however, omitted in this paper due to space limitations. 

Note that for $K\geq5$ the numerical algorithm outlined~\cite{2003Dighe,2005Maaref} needs to be used to obtain the coefficients. Interested readers may contact the first author for a copy of the code of the numerical algorithm implemented in Mathematica\textregistered.

\section{Numerical Results}\label{sec:simu}

\begin{figure}[t!]
\centering
\includegraphics[width=3.5in]{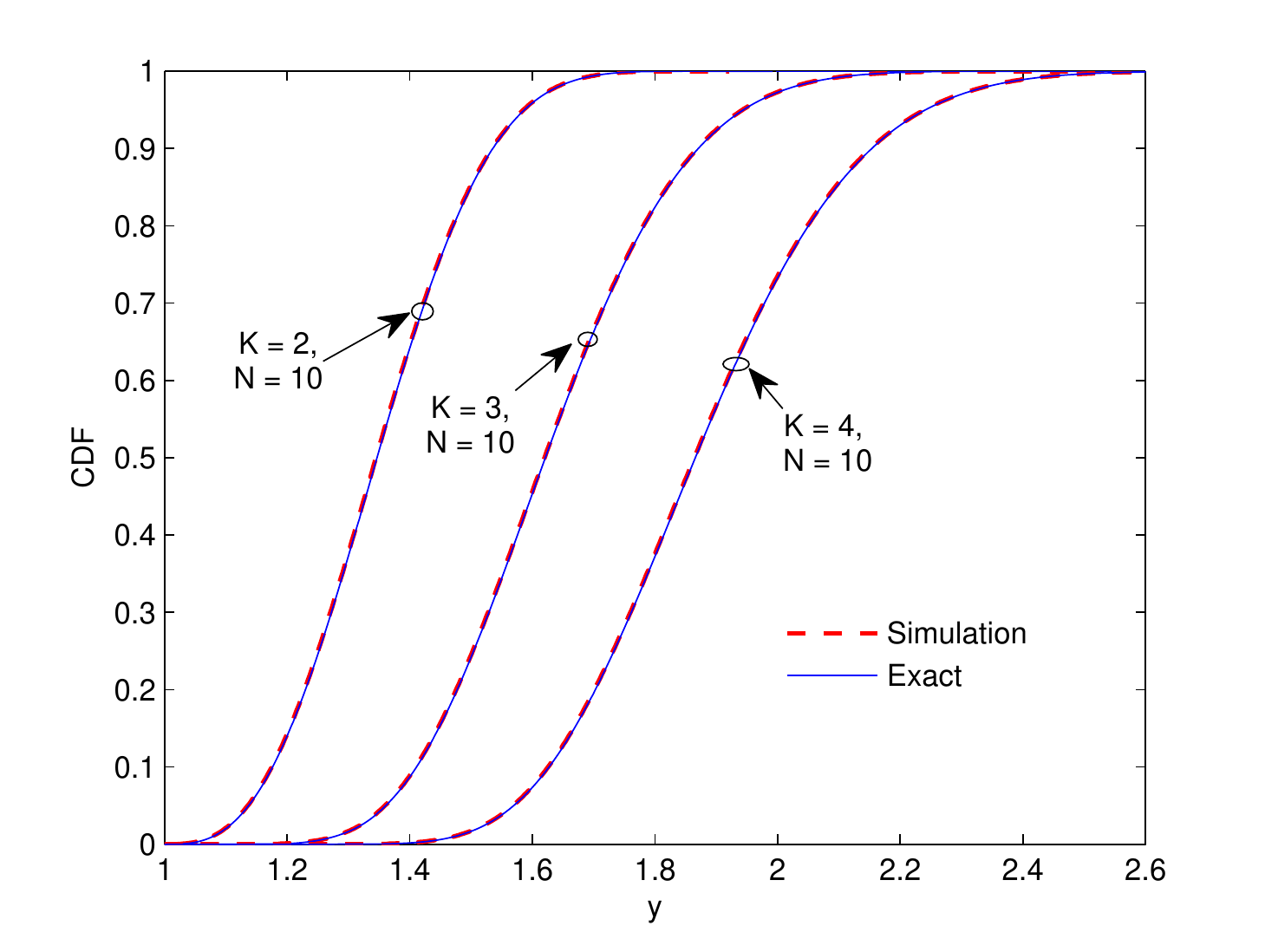}
\caption{CDF of the scaled largest eigenvalue using closed-form coefficients.}\label{fig:2}
\end{figure}

\begin{figure}[t!]
\centering
\includegraphics[width=3.5in]{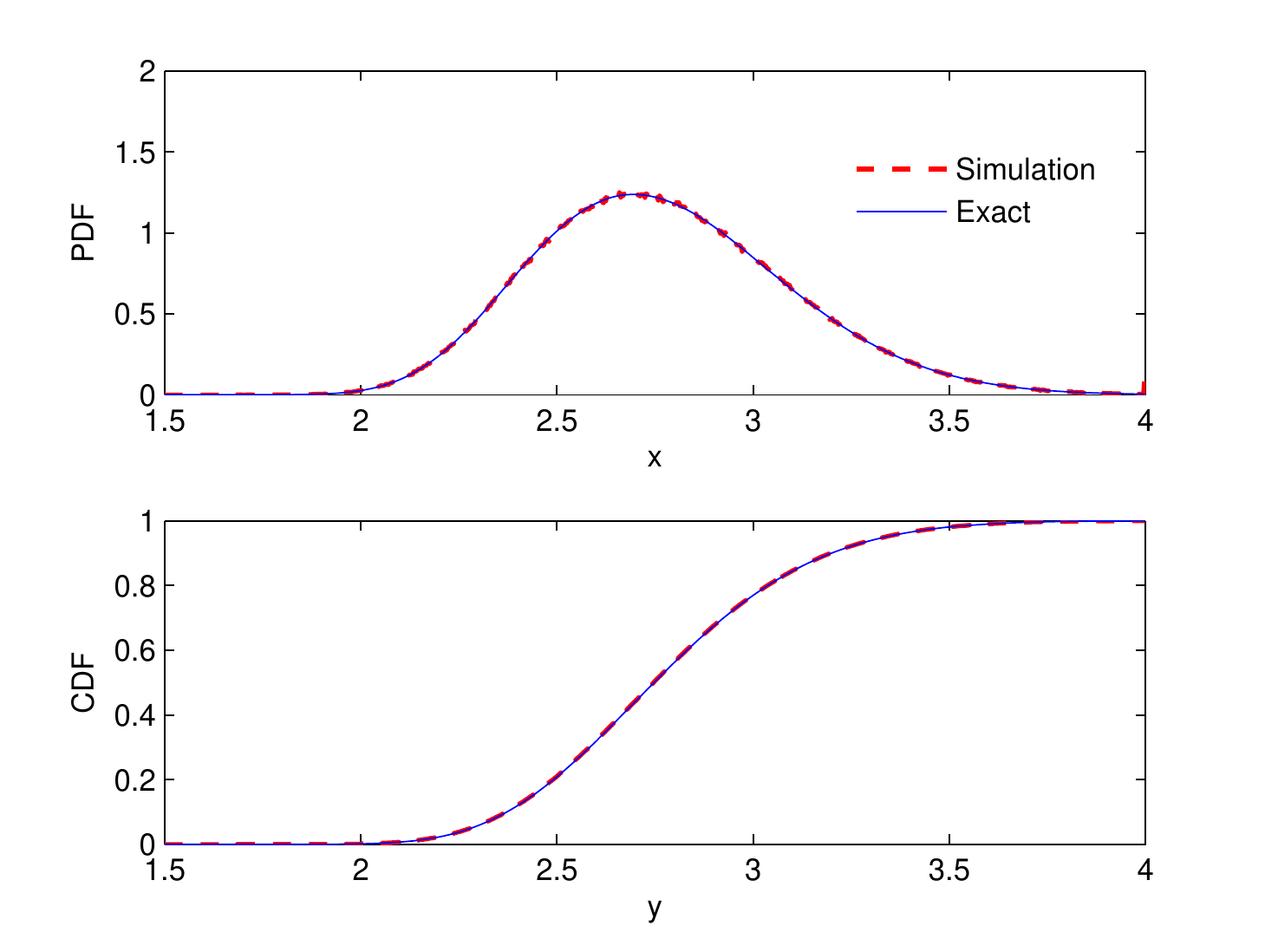}
\caption{PDF and CDF of the scaled largest eigenvalue: $K=6$, $N=6$, using numerical coefficients.}\label{fig:3}
\end{figure}

Some numerical examples are provided in this section to verify the derived SLE PDF expression~(\ref{eq:fXsimp}), CDF expression~(\ref{eq:FXsimp}) and corresponding coefficient expressions. We first examine the cases when both closed-form distribution and coefficients are available, where we choose $N=10$ with $K=2, 3, 4$ respectively. Using the derived CDF expression~(\ref{eq:FXsimp}) with the corresponding closed-form coefficients, the CDFs against simulations are plotted in Fig.~\ref{fig:2}. In addition, we consider the case when $K=6$ and $N=6$, where the closed-form coefficients are not available. In this case, inserting these numerically obtained coefficients from~\cite{2003Dighe} (Table~$\Rmnum{4}$) into~(\ref{eq:fXsimp}) and~(\ref{eq:FXsimp}), the corresponding PDF and CDF are drawn in Fig.~\ref{fig:3}. From Fig.~\ref{fig:2} and Fig.~\ref{fig:3}, we can see that the derived results match simulations well.

\section{Conclusions}\label{sec:conc}

Knowledge on the statistical property of the scaled largest eigenvalue of Wishart matrices is key to understanding the performance of various hypothesis testing procedures and communication systems. In this work, we derived exact expressions for the PDF and CDF of the SLE for arbitrary matrix dimensions by using a Mellin transform based method. Our results are easy and efficient to compute, they do not involve any complicated integral representations or unexplicit expressions as opposed to existing results. The derived expressions were validated through simulation results.

\section*{Acknowledgment}

This first author wishes to thank Shang Li from HKUST for his discussions on this topic.

\ifCLASSOPTIONcaptionsoff
\newpage
\fi

%\IEEEtriggeratref{2}

\end{document}